\documentstyle[12pt]{article}

\newcommand{\bce}{\begin{center}}
\newcommand{\ece}{\end{center}}
\newcommand{\beq}{\begin{equation}}
\newcommand{\eeq}{\end{equation}}
\newcommand{\bea}{\vspace{0.25cm}\begin{eqnarray}}
\newcommand{\eea}{\end{eqnarray}}

\newcommand{\ba}{\begin{array}}
\newcommand{\ea}{\end{array}}

\newcommand{\doublespace}{
    \renewcommand{\baselinestretch}{1.6}\large\normalsize}

\def\lsim{\mathrel{\rlap{\lower4pt\hbox{\hskip1pt$\sim$}}
    \raise1pt\hbox{$<$}}}         
\def\gsim{\mathrel{\rlap{\lower4pt\hbox{\hskip1pt$\sim$}}
    \raise1pt\hbox{$>$}}}         

\def\beq{\begin{equation}}
\def\endeq{\end{equation}}
\def\arr{\begin{eqnarray}}
\def\endarr{\end{eqnarray}}
\makeindex

  \advance\oddsidemargin  by -1in
  \advance\evensidemargin by -1in
  \advance\topmargin      by -0.5in
\begin{document}
 
 
\begin{center}
{\Large \bf Scaling properties of transverse flow in Bjorken's 
scenario for heavy ion collisions
\vspace{1.0cm}\\}
{\large \bf V.Fortov, P.~Milyutin $^{a}$ and
N.~Nikolaev$^{b,c}$ \vspace{1.0cm}\\}
{\it
$^{a}$High Energy Density Research Center of the Russian Academy of
Sciences, IVTAN, Izhorskaya 13/9, 127412 Moscow, Russia
\medskip\\
$^{b}$Institut f. Kernphysik, Forschungszentrum J\"ulich, D-52425 J\"ulich,
Germany
\medskip\\
$^{b}$L. D. Landau Institute for Theoretical Physics, GSP-1,
117940, \\
ul. Kosygina 2, Moscow V-334, Russia.\vspace{1.0cm}\\ }
{\Large \bf
Abstract}\\
\end{center}
We report a simple analytic solution for the velocity $u$ of the
transverse flow  of QGP at a hadronization front 
in Bjorken's scenario. We establish scaling properties of the transverse
flow as a function of the expansion time. We present simple scaling 
formula for the expansion velocity distribution.
\bigskip\\

 \doublespace
\pagebreak
 
Landau's hydrodynamic stage \cite{Landau} is a 
part of all scenarios for the evolution of the hot and dense matter 
(quark-gluon plasma - QGP) formed in
ultrarelativistic heavy ion collisions \cite{Reviews}. It is now well 
understood that Landau's complete stopping of Lorentz-contracted colliding 
nuclei is not feasible because of the Landau-Pomeranchuck-Migdal (LPM) 
effect, i.e., the finite proper formation time $\tau_{0}$ (\cite{LPM}, 
for the modern modern QCD approach to the LPM effect see \cite{BGZ}, the
early works on LPM phenomenology of nuclear collisions are reviewed in 
\cite{NNN}), although evaluations of $\tau_{0}$ and of the initial 
energy density $\epsilon_{max}$ remain 
controversial \cite{Reviews}. The corollary of the LPM effect 
in conjunction with the approximate central rapidity plateau is the 
rapidity-boost invariance of initial conditions. The corresponding 
solution for a longitudinal expansion in an 1$+$1-dimensional approximation, 
neglecting  the transverse flow, was found by Bjorken (\cite{Bjorken}, 
see also \cite{Shuryak}). There is some experimental evidence 
\cite{Bearden,Appelshauser}, although a disputed one 
\cite{Stachel,Cleymans}, for a transverse flow  
which must develop if the lifetime of the hydrodynamical stage is 
sufficiently long. 

In this communication we present a simple solution of the Euler-Landau 
equation for the velocity of transverse expansion, $u,$ gained in the 
hydrodynamic expansion of QGP  
before the hadronization phase transition. Our 
solution shows that for the usually considered lifetime $\tau_{B}$
of QGP the transverse flow is non-relativistic. It is only marginally 
sensitive to properties of the hot stage and offers a reliable 
determination of $\tau_{B}$ if the radial profile of the initial energy 
density is known. We find that the $u$-distribution is a scaling function 
of $u/u_{m}$, where $u_{m}$ is a maximal velocity of expansion. 

We start with the familiar Landau relativistic hydrodynamics equations 
\beq
\partial_{\mu}T_{\mu\nu}=0 \, ,
\label{eq:1}
\endeq
for the energy-momentum tensor $T_{\mu\nu}=(\epsilon + p)u_{\mu}u_{\nu}-
p\delta_{\mu\nu}$, where $\epsilon$ and $p$ are the energy density and 
pressure in the comoving frame, and $u_{\mu}$ is the 4-velocity of the 
element of the fluid \cite{Landau,LandauLifshitz}. The initial state is 
formed from subcollisions of constituents 
(nucleons, constituent quarks and/or partons) of colliding nuclei and 
is glue dominated at early stages. The LPM effect implies that for a 
subcollision at the origin, $x=(t,z,\vec{r})=0$, the secondary particle 
formation vertices lie on a hyperbole of constant proper time $\tau$,
$
\tau^{2}=t^{2}-z^{2}\sim \tau_{0}^{2} \, ,
$
and $\epsilon,p$ do not depend on the space-time rapidity
$
\eta = {1\over 2} ln\left( {t+z \over t-z} \right)
$
of the comoving reference frame \cite{Bjorken,Shuryak}. In the 
1$+$1-dimensional approximation, this leads to the celebrated Bjorken
equation
\beq
{\partial\epsilon \over \partial\tau}+
{\epsilon + p \over \tau}=0  \, .
\label{eq:2}
\endeq
According to the lattice QCD studies, the familiar 
\cite{LandauLifshitz} $c_{s}^{2}={1\over 3}$ holds for a velocity
of sound $c_{s}$ in the QGP excepting a negligible narrow 
region of the hadronization transition temperature $T_{h} \sim 160 $MeV 
and energy density $\epsilon_{h} \sim 1.5$ GeV$/$fm$^{3}$ \cite{Hung,Schlei}. 
With the equation of state $p=c_{s}^{2}\epsilon$, the Bjorken equation
has a solution 
\beq
\epsilon \propto \tau^{-(1+c_{s}^{2})}
\label{eq:3} 
\endeq
Widely varying estimates for $\epsilon_{max}$ and the proper time 
$\tau_{0}$ are found in the literature \cite{Reviews,Schlei,Eskola,Heinz}. 
However, as Bjorken has argued \cite{Bjorken}, $\epsilon_{max} \propto 
{1\over \tau_{0}}$ and the actual dependence of the Bjorken lifetime 
$\tau_{B}$ on $\epsilon_{max}$ is rather weak:
\beq
\tau_{B} = \tau_{0}
\left[\left({T_{max} \over T_{h}}\right)^{4 \over 1+c_{s}^{2}}-1\right]
={\tau_{0} \epsilon_{max} \over \epsilon_{h}} 
\left[\left({\epsilon_{max} \over \epsilon_{h}}\right)
^{c_{s}^{2}\over 1+c_{s}^{2}}-{\epsilon_{h} \over \epsilon_{max}}\right]\,.
\label{eq:4}
\endeq
For central $ PbPb$ collisions, for which there is some experimental 
evidence for the QGP formation \cite{Reviews}, the typical estimates 
are $\tau_{B}\approx 3 {\rm f/c}$ at SPS \cite{Schlei,Dumitru} and 
$\tau_{B} \approx 6 {\rm f/c}$ at  RHIC \cite{Reviews}, which are
much larger than the standard estimate $\tau_{0}\sim 0.5$ f/c. 

Now we turn to the major theme of collective transverse expansion, 
which is driven by radial gradient of pressure. As we shall see,
the radial flow is nonrelativistic. Then, to the first order in 
radial velocity $u_{r}$, the radial projection of (\ref{eq:1}) 
gives the Euler-Landau equation
\beq
(\epsilon + p){\partial u_{r} \over \partial \tau }+
u_{r}\left(
{\partial(\epsilon + p) \over \partial \tau }+
{(\epsilon + p) \over \tau }\right)+ {\partial p \over \partial r}
=0\, ,
\label{eq:7}
\endeq
in which we can use the Bjorken's solution for $\epsilon$ and 
$p=c_{s}^{2}\epsilon$. Then the Euler-Landau equation can be cast in a 
simple form
\beq
{\partial u_{r} \over \partial \tau }-{c_{s}^{2} \over \tau}
u_{r}= -{c_{s}^{2} \over( 1+c_{s}^{2})}{\partial \log p \over \partial r}\,.
\label{eq:8}
\endeq
The important point is that the transverse expansion of the QGP fireball 
can be neglected which we can justify {\it a posteriori}. For this reason 
the time dependence of the logarithmic derivative $D(r,\tau)={\partial 
\log p \over \partial r}$ can be neglected, it is completely determined
by the initial density profile and depends neither on the temperature nor 
fugacities of quarks and gluons, which  substantially reduces the 
model-dependence of the transverse velocity. Then the  solution of 
(\ref{eq:7}) subject to the boundary condition 
$u_{r}(\tau_{0})=0$ is 
\beq
u_{r}(r,\tau)= 
{c_{s}^{2} \tau^{c_{s}^{2}}
\over 1-c_{s}^{4}} \int_{\tau_{0}}^{\tau} dt { t^{-c_{s}^{2}}
D(r,t)} \approx
{c_{s}^{2}\tau D(r,0)\over 1-c_{s}^{4}}
\cdot
\left[1 - \left({\tau_{0} \over \tau}
\right)^{1-c_{s}^{2}}
\right] \, .
\label{eq:9}
\endeq

The model-independent estimates for the initial density/pressure profile 
are as yet lacking. AT RHIC and higher energies of the initial state is 
expected to be formed by semihard parton-parton interactions for which 
nuclear shadowing effects can be neglected \cite{Reviews,Eskola}. Then 
for central collisions $\epsilon(r,\tau_{0}) \propto T_{A}^{k}(r)$, 
where $k=2$ and $T_{A}(r)=\int dz n_{A}(\sqrt{z^{2}+r^{2}})$ is the 
density of constituents, $T_{A}(r) \sim \exp(-{r^{2}\over R_{A}^{2}})$,
where $R_{A}\approx 1.1A^{-1/3}$fm is the nuclear radius. In another 
extreme scenario of strong shadowing and of strong LPM effect
the soft particle production is not proportional to the multiplicity 
of collisions of fast partons \cite{BGZ,NNN} and $k=1$ is more appropriate. 
Hereafter we take $k=2$. In any case, the logarithmic pressure gradient
is approximately linear, $D(r,t)\approx 2kr/R_{A}^{2}$, and according
to the solution (\ref{eq:9}) the displacement of the fluid element
$\Delta r$ is proportional to the radius, $\Delta r \propto r\tau^{2}$. 
Consequently, we have the Hubble-type radial rescaling 
\beq
\lambda(\tau)\approx 1 +{\Delta r \over r} \approx 
1 + \left({\tau\over \tau_{T}}\right)^{2} \, ,
\label{eq:10}
\endeq
where 
\beq
\tau_{T}\approx {R_{A}\over \sqrt{k}c_{s}} \, .
\label{eq:11}
\endeq
has a meaning of the {\sl lifetime against transverse expansion}. 
For central $ PbPb$ collisions eq.~(\ref{eq:11}) gives $\tau_{T} \approx 
10 k^{-0.5} {\rm f/c}$ which is larger than the above cited estimates 
of $\tau_{B}$ and at SPS and RHIC energies the transverse expansion of 
the fireball can be neglected. 

In a quasi-uniform plasma  the hydrodynamic expansion lasts until 
$\epsilon = \epsilon_{h}$.
For long-lived QGP and $\tau \gg \tau_{0}$ we can use the Bjorken's solution
\beq
\epsilon(r,\tau)=\epsilon_{h}\left({T_{A}(r)\over T_{A}(0)}\right)^{k}
\left({\tau_{B} \over \tau}\right)^{1+c_{s}^{2}}\, ,
\label{eq:11*}
\endeq
which gives the position of the hadronization front 
\beq
r_{h}(\tau)=R_{A} \sqrt{{1+c_{s}^{2}\over k} \ln{\tau_{B}\over \tau}}
\label{eq:12}
\endeq
and the radial velocity at the hadronization front 
\beq
u(\tau)=u_{r}(r_{h}(\tau),\tau)={c_{s}^{2}\tau \over R_{A}(1-c_{s}^{4})}
\sqrt{4k(1+c_{s}^{2})\ln{\tau_{B}\over \tau}} 
\left[1-\left({\tau_{0}\over \tau}\right)^{1-c_{s}^{2}}\right]\,.
\label{eq:13}
\endeq    
For the usually discussed $\tau_{B}$ and $\tau_{0}$ we have $\tau_{B}\gg 
\tau_{0}$. For such s long-lived QGP,  $\tau_{B} \gg \tau_{0}$, the radial 
velocity takes the maximal value $u_{m}$ at $t\approx 1/\sqrt{e}$, 
\beq
u_{m}={c_{s}^{2}\tau_{B}\over R_{A} (1-c_{s}^{4})} 
\sqrt{2k(1+c_{s}^{2})\over e}
\left[1-\left({\tau_{0}\sqrt{e}\over \tau_{B}}\right)^{1-c_{s}^{2}}\right]\,.
\label{eq:14}
\endeq
It is remarkable that the average radial acceleration $u_{m}/\tau_{B}$ 
is approximately constant. The solid line in fig.~1 shows the maximal 
velocity $u_{m}$ evaluated from (\ref{eq:13}). The large-$\tau_{B}$ 
approximation (\ref{eq:14}), shown by the dashed line, reproduces these 
results to better than $\sim 4\%$ at $\tau_{B}=3$ f/c and better than 
$\sim 1\%$ at $\tau_{B}=8$ f/c. For the above cited estimates for $\tau_{B}$ 
in central $PbPb$ collisions we find $u_{m}(SPS)\approx 0.13$  and 
$u_{m}(RHIC)\approx 0.28$, consequently the nonrelativistic expansion
approximation is justified very well. Now notice, that for a long-lived 
QGP the hadronization front (\ref{eq:12}), shown in fig.~2, and 
$u(\tau)/u_{m}$ depend only on the scaling variable $t=\tau/\tau_{B}$,
with an obvious exception of the short-time region $\tau\sim \tau_{0}$. 
This scaling property is clearly seen in fig.~3 where we show the 
$u/u_{m}$ as a function of $t=\tau/\tau_{B}$. Notice a convergence to 
a universal curve with the increasing $\tau_{B}$. 

The most interesting quantity is a radial velocity distribution which
can be evaluated experimentally from the Doppler modifications of the 
thermal spectrum. In order to test our results one needs particles which 
are radiated from the hadronization front. The standard scenario is 
that the hadronization transition is followed by an expanding mixed 
phase which, however, does not contribute to the transverse velocity 
because in the mixed phase $c_{s}^{2}0$ is negligible small. 
The mixed phase is followed 
by a hydrodynamic expansion and post-acceleration of strongly interacting 
pions and baryons until the hadronic freeze-out temperature $T_{f}< T_{h}$ 
is reached \cite{Leutwyler}. This post-acceleration is negligible for 
weakly interacting $K^{+}$ and $\phi$-mesons, which gives the desired
access to the radial flow at the hadronization 
transition. In the evaluations of the modification of the the thermal 
spectrum and one needs to know the $u$-distribution weighted with the 
particle multiplicity. For the of constant hadronization temperature 
contribution of the hadronization surface $r_{h}(\tau)$ to the particle 
multiplicity is  
\beq
dw \propto r_{h}(\tau)d\tau\,.
\label{eq:14*}
\endeq
Making use of the solutions (\ref{eq:12}) and (\ref{eq:13}), it can 
readily be transformed into $dw/du \propto r_{h}(du/d\tau)^{-1}$.
The important point is that because of the above discussed scaling 
properties of the transverse flow, the velocity distribution is a 
scaling function of $x=u/u_{m}$:
\beq
{dw \over du}={1\over u_{m}}{f(x,\tau_{B})\over \sqrt{1-x}}\, ,
\label{eq:15}
\endeq  
where for a long-lived QGP $f(x,\tau_{B})$ does not depend on $\tau_{B}$.
The square-root singularity at $x=1$ is a trivial consequence of the 
vanishing derivative $du(\tau)/d\tau$ at $\tau \approx \tau_{B}/\sqrt{e}$.
In Fig.~4 we show the scaling function $f(x,\tau_{B})$ for $\tau_{B}=6$ f/c.
We don't show  $f(x,\tau_{B})$ for other values of $\tau_{B}$, because
the variations from $\tau_{B}=6$ f/c to 3 f/c to 9 f/c do not exceed 
several per cent and are confined to a narrow region of  $x\lsim 0.2$.
The approximation $f(x)=0.5$ is good for all the practical purposes.

In conclusion, we would like to argue that the shape of the velocity 
distribution is to a large extent the model independent one. The generic
origin of the  square-root singularity at $x=1$ has already been 
emphasized, the fact that $f(0)\neq 0$ is due to a radiation from the 
surface $r_{h}\sim R_{A}$ at early stages of expansion.  
Above we assumed 
that hydrodynamic expansion continues untill the hadronization 
transition. Following Pomeranchuk \cite{Pomeranchuk} one can argue 
that in the non-uniform plasma the hydrodynamic expansion stops 
when the mean free path 
\beq
l_{int}={1 \over n(r_{c},\tau)\sigma_{t}}
\label{eq:24}
\endeq
defined in terms of the transport cross section $\sigma_{t}$, is 
larger that the GQP density variation length $D(r,0)^{-1}$. In the 
partially equilibrated QGP $l_{int} \propto T^{-1}$. Then the 
Pomeranchuk condition gives the temperature $T_{P}$ at which the 
hydrodynamic expansion stops, $T_{P} \propto r/R_{A}^{2}$. The 
possibility remains open that at early stages $T_{P} > T_{h}$, 
in which case $dw \propto r_{h}(\tau)(T_{P}/T_{h})^{3}d\tau$. 
This enhanced radiation at early stages at slow radial expansion
but at higher temperatures $T_{P}$ would mimic radiation at a lower
temperature and higher radial velocity. This may result in the
apparent depletion of $f(0)$; in order to explore 
this possibility one needs better understanding of $l_{int}$ 
near the hadronization transition.
 
The NA49 fits to the proton, kaon and pion transverse mass $m_{T}$ 
distribution in central $PbPb$ collisions at SPS assume identical 
freeze-out temperature for all particle species \cite{Bearden}. For 
positive particles NA49 finds $T_{f}= 140\pm 7$ MeV and the transverse 
velocity $\langle u \rangle = 0.41\pm 0.11$. However, for the $K^{+}$ 
one must take the higher freeze-out temperature $T_{f}=T_{h}\approx 160$
MeV given by the lattice QCD. Because of the anti-correlation between the 
local temperature $T_{f}$ and $\langle u_{T}\rangle$, see Fig.~7 in 
\cite{Appelshauser}, such a fit with larger $T_{f}$ to the same $m_{T}$ 
distribution shall yield smaller $\langle u \rangle$.
 
{\bf Acknowledgements:}
This work was partly supported by the INTAS grant 96-597
and the Grant N 94-02-05203 from the Russian Fund for Fundamental
Research.

\pagebreak

\pagebreak
{\bf \Large Figure captions}
\begin{itemize}
\item[Fig.~1] 
The maximal velocity of radial expansion $u_{m}$ for for central $PbPb$
collisions as a function 
of the expansion time $\tau_{B}$. Shown by the dotted line is
the large-$\tau_{B}$ formula (\ref{eq:15}). 

\item[Fig.~2] 
The time dependence of the hadronization front for central $PbPb$
collisions.

\item[Fig.~3] 
The converegence to the scaling behaviour of the time dependence of 
the relative radial velocity $u/u_{m}$  
for central $PbPb$ collisions. 

\item[Fig.~4]
The scaling function $f(x,\tau_{B})$ of. Eq.~(\ref{eq:15}) is shown for 
$\tau_{B}=6$.

\end{itemize}
\end{document}